\documentclass[useAMS,usenatbib]{mn2e}
\usepackage{graphicx}  
\usepackage{epsf}
\usepackage{amsmath}
\newcommand{\hangpar}{\noindent\hangindent.1in}
\newcommand{\teff}[1]{${T_{\rm eff}}$}
\newcommand{\logg}[1]{${\log(g)}$}
\newcommand{\vsini}[1]{$v\cdot\sin(i)$}
\def\cm1{$\rm cm^{-1}$}
\def\kms{$\rm km\,s^{-1}$}
\def\DE{D\kern-0.75em \raisebox{1.0pt}{=}\ }

\def\Sum{N_{\rm tot}}
\def\I{{\sc i}}
\def\II{{\sc ii}}
\title[Chemical abundances of the high-latitude Herbig Ae
star PDS2]
{Chemical abundances of the high-latitude Herbig Ae Star PDS2
}
\author[C. R. Cowley,
S. Hubrig \& N. Przybilla]
{C. R. Cowley${^1}$
\thanks{E-mail: cowley@umich.edu},
S. Hubrig${^2}$,
and N. Przybilla${^3}$
 \\
$^{1}$Department of Astronomy, University of Michigan,
   Ann Arbor, MI 48109-1042, USA\\
$^{2}$Leibniz-Institut f\"{u}r Astrophysik Potsdam (AIP), 
An der Sternwarte 16, 14482, Potsdam, Germany  \\
$^{3}$Institute f\"{u}r Astro- und Teilchen Physik, Technikerstr. 25/8, A-6020 Innsbruck, Austria\\
}
\begin{document}

\date{Accepted . Received ; in original form }

\pagerange{\pageref{firstpage}--
\pageref{lastpage}} \pubyear{2013}

\maketitle

\label{firstpage}

\begin{abstract}
The Herbig Ae star PDS2 (CD $-$53$^\circ$ 251) is
unusual in several ways.  It has a high Galactic
latitude, unrelated to any known star-forming region.
It is at the cool end of the Herbig Ae sequence, where
favorable circumstances facilitate the determination of
stellar parameters and chemical abundances.  We find
$T_{\rm eff} = 6500$ K, and $\log(g) = 3.5$. The
relatively low $v\cdot\sin(i) = 12\pm2$ \kms\, made it
possible to use mostly weak lines for the abundances.
PDS2 appears to belong to the class of Herbig Ae stars
with normal volatile and depleted involatile elements,
thus resembling the $\lambda$ Boo stars.  The
intermediate volatile zinc consistently violates this
pattern. 
\end{abstract}
\begin{keywords}
--stars:Herbig Ae
--stars:abundances
--stars:individual: PDS2
--stars:individual: HD 104237   
--stars:individual: HD 101412   
--stars:individual: HD 190073
\end{keywords}
\section{Introduction}
\label{sec:intro}
Gregorio-Hetem, et al. (1992) describe their survey of
stars in the IRAS Point Source Catalog (1988) for T Tauri
stars. Their work and follow-up studies became known as
the Pico dos Dias Survey (PDS).  It located a number of Herbig
Ae/Be candidates, among them the isolated, high-latitude
object, PDS2 (CD $-$53 251). Subsequent studies (e.g. Vieira, et al.
2003, Pogodin et al. 2012) confirmed the classification
as a young stellar object.
The $\delta$ Scuti-like pulsations discovered by 
Bernabei, et al. (2007) were confirmed by Marconi et al.
(2010), who noted that it could ``constrain the poorly
sampled red edge of the PMS [pre main sequence] instability strip.''  
The presence of a weak magnetic field in this star is
uncertain, as the star was observed only on two different
epochs (Wade, et al. 2007, Hubrig et al. 2009).  Bagnulo,
et al. (2012) conclude that a field is possible, but 
``certainly not yet definitely established.''


The metallic-line (absorption) spectrum of PDS2 is well developed, as its
temperature places it among the mid F-type stars. Emission is restricted
to the hydrogen and helium lines and [O I].

The PDS2 spectrum has relatively sharp lines, which enables
us to use weak lines whose equivalent widths are independent of
instrumental or rotational broadening.  This is not the case for
many Herbig Ae stars.

The youth of PDS2 and its high Galactic latitude pose the question
of whether the star was born close the the Galactic plane.  The
available kinematical data are not precise enough to answer this
question unequivocally.

This paper may be considered a sequel to three previous abundance
studies of the Herbig Ae stars HD 101412, HD 190073, and HD 104237
(Cowley et al. 2010, Cowley \& Hubrig 2012, and Cowley et al. 2013).
We refer to them as Papers I, II, and III, respectively.

\section{Observational material}
\label{sec:spec}

\begin{center}
\begin{table}
\caption{Observations.  See text for explanation and references.\label{tab:epoch}}
\begin{tabular}{l l r l} \hline
Spectrum & epoch(JD245) & S/N &$\lambda/\Delta\lambda$ \\ \hline
HARPS &4781.08/4782.04 &$\sim$45&$\sim$60000  \\
X-shooter&5375.5&550 &$\sim$17400  \\
CRIRES&6153.5/6172.5&175&$\sim$90000 \\  \hline
\end{tabular}
\end{table}
\end{center}

The observational materials are summarized in
Table~\ref{tab:epoch}.  Note that the entries
for signal to noise (S/N), and 
resolving power (RP or $\lambda/\Delta\lambda$)
are typical values.  Individual measurements can vary by as
much as 30\%. Resolution and S/N for X-shooter
are from Pogodin et al. (2012), for CRIRES from
Cowley, et al. (2012).

Equivalent width measurements were made on 9 
averaged HARPS (Mayor, et al. 2003)
spectra downloaded from the ESO archive.  
We
examined two sets of 9 spectra obtained on the
nights of 11/12 November and 13 November 2008 UT.
The first set of spectra were clearly of higher
quality than the second, and all measurements 
were therefore based on that set.  The largest
difference in barycentric radial velocities
found in the fits headers for
the adopted spectra was 0.079 \kms.  
We attempted
to allow for the differences in radial velocity,
but the results were no better than straight
averages, which we adopted.  

To find the signal-to-noise ratios, we fit 
second-order curves to portions of the spectra
that appeared to be free of absorption lines,
or contained the smooth wing of a strong line.
The S/N is taken to be the standard deviation
of points from these fits.

\begin{figure}
\includegraphics[width=75mm,height=85mm,angle=-90]{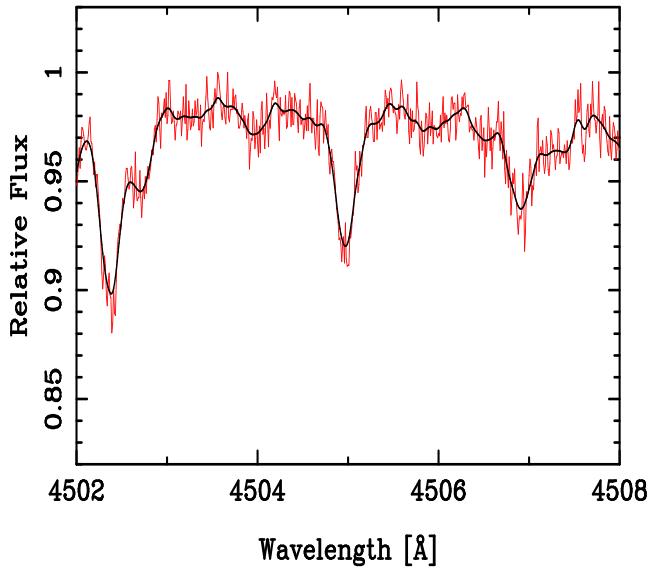}
 \caption{Sample region showing averaged (gray/red online) 
and Fourier filtered (black) spectrum.
\label{fig:plfint}}
\end{figure}

The added spectra were still noisy, and were Fourier
filtered using the standard Brault-White (1971)
algorithm.  Typical values of S/N for one of the
individual exposures are $\sim$45.  After addition
the S/N increases to typically $\sim$130. Filtering
further reduces the unwanted, high-frequency power by
a factor of $\sim$2.5, to $\sim$320, as
illustrated in Fig. \ref{fig:plfint}.   The
latter value is strictly not S/N as the filtered
points are correlated.

The resolution of the averaged and filtered spectrum is
about 60,000, based on fits of gaussians to telluric
lines, and the sharp, stellar [O I] (see
Sec.~\ref{sec:spec2}).

For the O \I\, triplet, outside the HARPS coverage we
used an X-shooter  (Vernet, et al. 2011)
spectrum.  
A CRIRES (K\"{a}ufl, et al. 2004)
spectrum is used to show the He \I\, $\lambda$10830
emission.

\section{Normalization and equivalent widths}

Normalization of the spectrum is done with a code
written by CRC.  The spectrum is viewed in 80 \AA\
intervals, and high points are located approximately
with a cursor. In a second pass, the actual highest
points are found within 0.5 \AA\ of the initially
chosen points.  Finally, the continuum is chosen to
pass through these points using a weighted combination
of linear interpolation and a cubic spline fit.

It is generally known that the wide profiles of the
Balmer lines are difficult to obtain from echelles,
such as HARPS or UVES.  Our normalization procedure
was a standard one. 
Perhaps the intrinsically narrower profiles
of a mid-F star makes the echelle order gap merging less
troublesome.  In any case, we have obtained good
agreement with the wing calculations and the observed
profiles of H$\beta$, H$\gamma$ (Figure
~\ref{fig:HgHarps}) and H$\delta$ from {\it both}
instruments (HARPS and X-shooter). With the exception
of the H$\alpha$ profile (Figure ~\ref{fig:HaUVX}), no
adjustments to the observed profiles were necessary to
achieve good agreement with the calculations.

Equivalent widths are measured individually, using
Voigt profile fits, as in previous studies (Papers I -
III).   The smallest equivalent width used in this
paper was 4.2 m\AA.  Features of the order of 1 or 2 m\AA\,
cannot be positively distinguished from noise.

\section{The spectrum of PDS2}
\label{sec:spec2}
Most of the spectrum of PDS2 is typical of a mid-F
star.  There are no metallic emission lines, which are
common in the spectra of young stars.  Emission at 
H$\alpha$ and He \I\, $\lambda$10830, in addition to the
infrared excess found in the IRAS survey are the 
basis of the assignment of PDS2 to the class of young
stars.

The rotational velocity was determined from the synthesis
of numerous stellar features (e.g. Mg \II\, $\lambda$4481).
We conclude this star has a relatively low $v\cdot\sin(i)$,
12$\pm$2 \kms.  This assumes that the macroturbulence has
a negligible half-width relative to the rotational half-width.
The $v\cdot\sin(i)$ could be lower if the macroturbulence 
were of comparable value.  The distinction is of little
consequence for abundances based almost exclusively on 
equivalent widths.

\begin{figure}
\includegraphics[width=75mm,height=85mm,angle=-90]{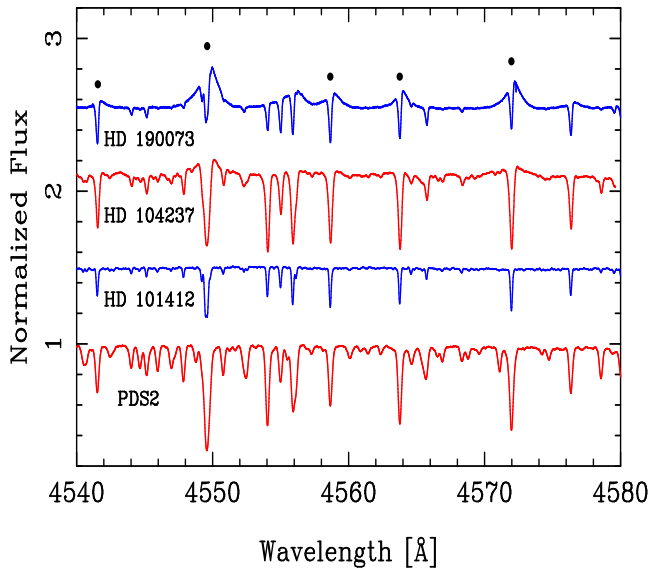}
 \caption{Sample spectra of four Herbig Ae stars. Filled
circles mark metallic lines sometimes in emission.  HD
190073 has the most developed metallic emissions,
followed by HD 104237 (DX Cha).  No emissions
are seen in this region of the HD 101412 spectrum,
or PDS2.  The top 3 spectra were discussed in Papers I,
II, and III.
\label{fig:pl4}}
\end{figure}

Figure ~\ref{fig:pl4}
compares a region of the spectra of 4 Herbig Ae
stars.  They are arranged in order of increasing
importance of metallic emission lines.
\begin{figure}
\includegraphics[width=75mm,height=85mm,angle=-90]{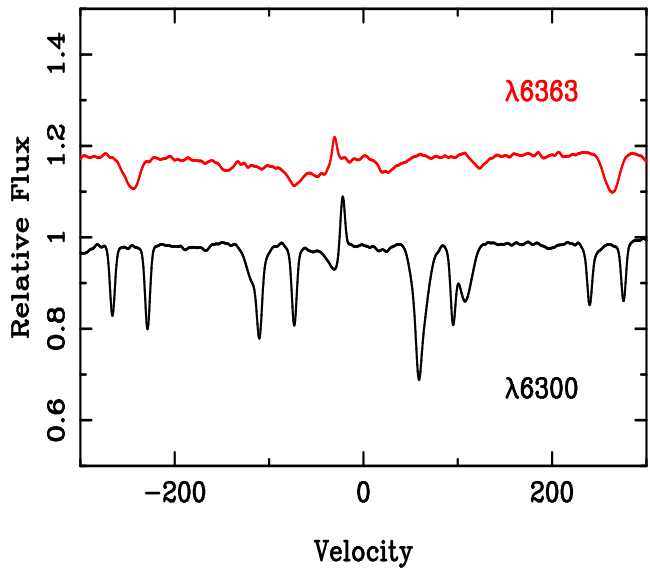}
 \caption{Emissions near [O \I] rest wavelengths
(Velocity = 0).  The $\lambda$6363
region (gray/red online) has been shifted upward for
purposes of display.  Both emissions are displaced to
the violet with respect to the stellar lines.  The
stronger line appears slightly less displaced
because of absorption on its violet edge.
\label{fig:forbid}} 
\end{figure}

Interestingly, sharp,
[O \I] lines are seen in emission, displaced
some 25-30 \kms\ to the violet  (Fig.~\ref{fig:forbid}).  
The violet portion of
the $\lambda$6300 line is absorbed by an atmospheric
and/or stellar feature, so an accurate velocity cannot
be determined.  Similar sharp [O \I] emissions were noted
in HD 190073 (Paper II) with a comparable
shift.  He \I, $\lambda$5876 is also weakly in emission.

\begin{figure}
\includegraphics[width=75mm,height=85mm,angle=-90]{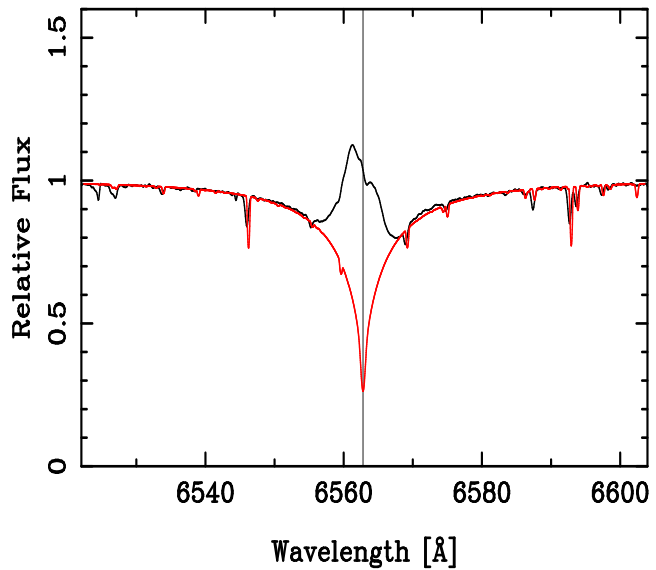}
\caption{H$\alpha$ from the X-shooter spectrum.  The
observed continuum (black) has been lowered by 1\% for
optimum fit  (gray/red online) to the line wings.
\label{fig:HaUVX}}
\end{figure}

\begin{figure}
\includegraphics[width=75mm,height=85mm,angle=-90]{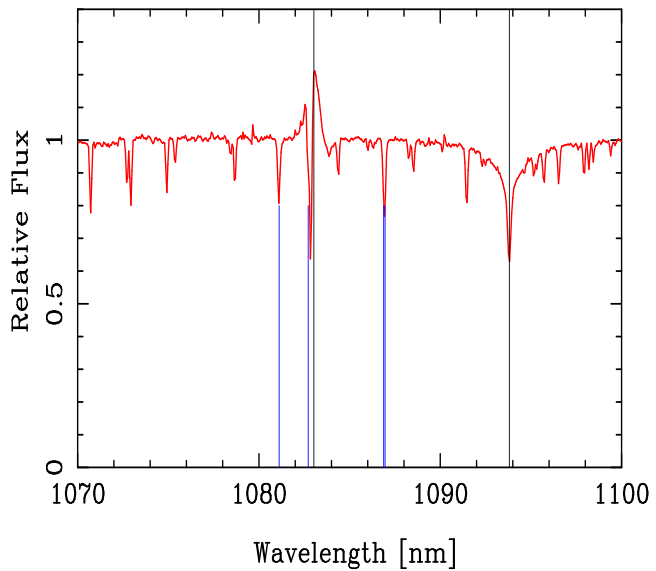}
 \caption{CRIRES spectrum showing He \I\
$\lambda$10830 in emission, and P$\gamma\, \lambda$10938
in absorption.  Vertical lines mark rest wavelengths;
 shorter lines are for Mg \I (10811.05\AA) and Si
\I (10827.09, 10868.79, and 10869.54\AA).  The remaining sharp absorptions are telluric or unidentified. 
\label{fig:plHe1}}
\end{figure}

\begin{figure}
\includegraphics[width=75mm,height=85mm,angle=-90]{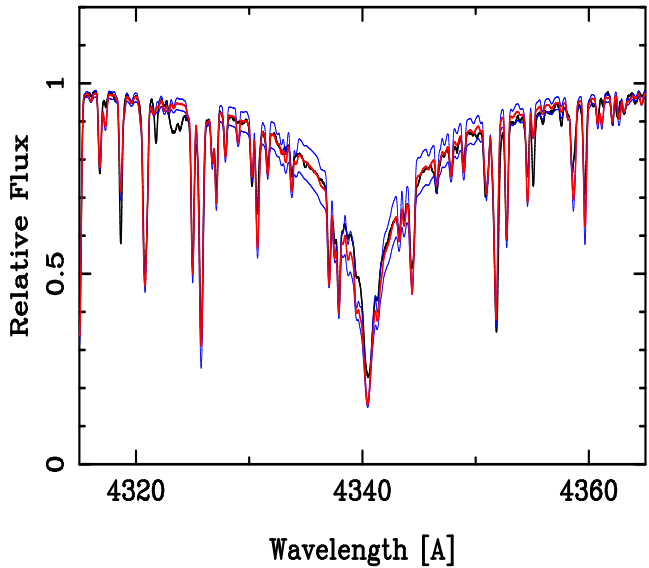}
 \caption{Averaged and filtered HARPS spectrum of
H$\gamma$ (black). The gray (red online) spectrum is
synthesized with the adopted model.  The thin lines
(blue online) show synthesized spectra with temperatures
$\pm$250 K from the adopted 6500 K model. Molecular features
were not included in the synthesis, which accounts for
some of the absorption not in the calculation, e.g. the
CH lines near $\lambda$4323-4325. \label{fig:HgHarps}}
\end{figure}

Apart from H$\alpha$ (Figure ~\ref{fig:HaUVX}) and He
\I\ $\lambda$10830 (Figure ~\ref{fig:plHe1}) there are
no strong emissions in our spectra of
PDS2.  Both lines are
variable.  See Hubrig, et al. (2013) for variability
of $\lambda$10830. Pogodin, et al. (2012) measured
equivalent widths at 8 phases for of H$\alpha$ and He
\I\, $\lambda$5876 from their X-shooter spectra.
Emissions at P$\beta$, P$\gamma$ and Br$\gamma$ were
measured at 5 phases.  These features were used to
determine mass accretion rates.  For PDS2, they found
$\log(\dot{M}_{\rm acc}) = -8.68$ in $M_\odot$
yr$^{-1}$.

\section{Atmospheric parameter determination}
\subsection{Model atmosphere and spectral codes} The model atmosphere code used here adopts the $T(\tau)$ relation from an ATLAS9 (Kurucz 1993, Castelli 
\& Kurucz 2003) calculation assuming solar abundances from the Castelli website\footnote{wwwuser.oat.ts.astro.it/castelli/}).  
The depth integrations are carried out by a code written at Michigan,
using (PDS2) abundances.  
We believe that the resulting minor inconsistency between the adopted $T(\tau)$ and $P(\tau)$ has no significant influence on the abundance determination presented below.

The atomic hydrogen and helium opacities are from Kurucz.    
For H$^-$, we used an interpolation formula in Gray (2005).
Topbase opacities are implemented for other elements, as described by Cowley, et al. (2003).  
These opacities are also used in the synthesis codes, which carry out flux integrations, which are then convolved with rotational profiles, as described, for example, by Gray (2005).  Equivalent width calculations are usually of one line only; occasionally a close-line pair will be synthesized, and the combined equivalent width used for an abundance.  Apart from oxygen, only  Co I, $\lambda$4867.9 was synthesized because of its location near H$\beta$.  The oxygen calculation is described separately below (Section \ref{sec:NLTE}).

The methods used here are the same as in Papers I-III.
Results from our codes have been verified by coauthors
using independent codes, and in independent studies
(e.g. Folsom, et al. 2012, 2013).

\subsection{Balmer lines}

It is well known that the strengths of the Balmer
lines are highly sensitive to the effective
temperatures of stars later than the early F's.  In
the temperature range of interest here, 6500 $\pm$
150K, there is at most a few per cent difference in
the profiles with \logg\ from 4.0 to 3.5, the latter
being the relevant range of surface gravities.

The H$\beta$ profiles of both the HARPS and X-shooter
spectra (not shown) have cores that are slightly less
deep than the calculations.  The violet edges of the
cores are also stronger than the calculated profiles.
The Balmer emission in PDS2 
is variable, as has been discussed, for example, by 
Pogodin, et al. (2012).

The H$\gamma$ and H$\delta$ profiles of PDS2 are hardly 
distinguishable from those of Procyon--apart from the 
stronger metallic lines in the latter star.  Most recent work
puts the effective temperature of Procyon between 6500
and 6600 K (Liebert, et al 2013, Boyajian, et al. 2013).
A calculated H$\gamma$ profile for PDS2 with 
$T_{\rm eff} =$ 6600 K
is slightly too strong, though possibly acceptable.  A
model with $T_{\rm eff} = 6700$ K is not acceptable.  If the 
temperature were as high as 6750 K, the abundance from
an Fe I line with a typical excitation of 2.2 eV would 
increase by 0.15 dex.

\subsection{Metal lines, gravity, turbulence, and abundances}

Generally, one may obtain a relation between the
effective temperature and surface gravity of a
star from the strengths of lines from neutral
and first-ionized atoms, e.g. Fe \I\ and \II.
In the temperature range that we have found for 
PDS2, the strengths of Fe \I\ lines are virtually
independent of surface gravity (see Gray 2005).  
If the lines are
weak, say $<$ 20 m\AA, their equivalent widths are
also independent of broadening mechanisms.  

We have used equivalent widths of {\it weak} Fe \I\ to
directly determine the iron abundance in PDS2. The
microturbulence ($\xi_t$) may then be determined from
intermediate and strong Fe \I\ lines.

Equivalent widths, abundances, wavelengths, and
excitation potentials for all atomic and ionic species
with more than 6 lines are read into a spreadsheet.
Plots are then made of the abundance vs. equivalent
width, excitation potential, and wavelength. The
microturbulence is chosen to minimize the dependence
of abundance on line strength.  We adopted the $\xi_t =
1.8$ km s$^{-1}$, first obtained for Fe I, the spectrum with
the most lines.  This value was compatible with other spectra
with numerous weak and strong 
lines, e.g. Cr \I, Cr \II, etc.  Because most of the lines used for
abundance were weak, results are not sensitive to this
parameter.

The surface gravity is then chosen to make abundances
from neutral and first-ionized species agree.

The weakest links in this chain of deduction for the 
atmospheric parameters are the absolute oscillator
strengths of the relevant species.  Our choices for the
oscillator strengths are given in Paper III  and
in the online material.

We tabulate abundances based on a model with $T_{\rm
eff} = 6500$ K.  The $\log(g) = 3.5$, but we estimate
a range of 3.75 - 3.3.  Most of the metal abundances,
based on the first spectra, are insensitive to gravity
uncertainties in this range.

\section{Abundances}
\label{sec:abundances}
\subsection{Oxygen (NLTE)\label{sec:NLTE}} 
\begin{figure}
\includegraphics[width=85mm,height=95mm,angle=-00]{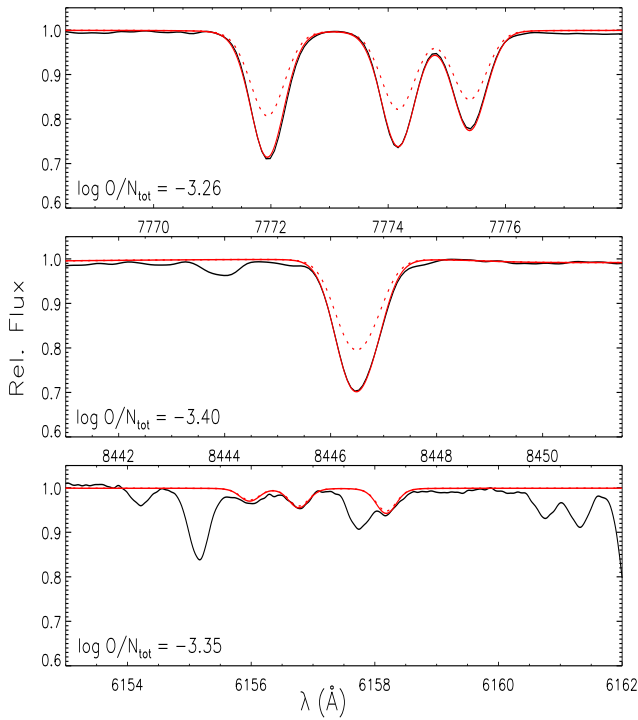}
 \caption{Spectrum synthesis of the observed O\,{\sc i} lines
(black), in non-LTE (full gray/red(online) lines) and LTE (dotted gray/red lines).
Best-fit abundances for the multiplets are indicated. Note that
only O\,{\sc i} $\lambda$6156.7\,{\AA} was considered for the
abundance determination in the lower panel, because of the blending of
the other components.
\label{fig:oxyg}}
\end{figure}

For the
determination of oxygen abundances from the strong
O\,{\sc i} $\lambda\lambda$7771-5\,{\AA} near-IR
triplet or from O\,{\sc i} $\lambda$8446\,{\AA},
non-LTE effects need to be accounted for. We employed
updated and extended versions of the {\sc
Detail/Surface} codes (Giddings 1981; Butler \&
Giddings 1985) for non-LTE line-formation computations
on the prescribed model atmosphere. The O\,{\sc i}
model atom of Przybilla et al. (2000) was used. This
was extended to account for collision strengths for
electron impact excitation by Barklem (2007),
excitation and ionization due to hydrogen collisions
employing the Steenbock \& Holweger (1984)
approximation (adopting a scaling factor $S_{\rm
H}$\,=\,1, determined from fitting the solar oxygen
spectrum).  Background opacities appropriate for
mid-F-type stars are from Przybilla \& Butler (2004),
with some improvements.  Line blocking is considered
via the LTE opacity sampling technique according to
Kurucz (1996).  The van der Waals broadening
coefficients are from Barklem et al. (2000).
Figure~\ref{fig:oxyg} shows the best fits to the
observed O\,{\sc i} multiplets. LTE line profiles for
the same abundances are indicated for comparison. The
non-LTE abundance corrections are essential, amounting
to about $-0.6$ to $-0.8$ \,dex for the stronger
lines. From the weak O\,{\sc i}
$\lambda\lambda$6155-8\,{\AA} triplet only the
component at 6156.7\,{\AA} was considered for the
abundance determination because of the strong blends
in the other lines. Blends of O\,{\sc i}
$\lambda$8446\,{\AA} with the weak Fe\,{\sc i}
$\lambda\lambda$8446.39/.57\,{\AA} lines were
accounted for, adopting our mean Fe\,{\sc i} abundance.

Oscillator strengths for $\lambda$7771-5 and 8446 are 
from Froese Fischer \& Tachiev (2004).  Those for
$\lambda$6155-8 are from Wiese, Fuhr \& Deters (1996).

\subsection{LTE for other elements\label{sec:LTE}}

Some studies have given NLTE corrections (cf. Hansen, et al.
2013), and for dwarfs near solar
abundances, they are rather small for Fe I, less than 0.1 dex (see
their Fig. 2).  A study by Mashonkina, et
al. (2011) discusses NLTE effects in
Procyon (Table 3).  Abundance results differ from LTE by
values under 0.03 dex, and might be zero, depending on
assumptions made about collisional rates with atomic
hydrogen.  An additional indication that LTE is a reasonable
approximation is that our abundances based on first and
second spectra are in good agreement, usually better than 
0.1 dex.

\subsection{Discussion of the results\label{sec:results}}


Table~\ref{tab:abtab} gives abundances in PDS2 and the
Sun as logarithms. 
The 50\% condensation temperatures
($T_c$), indicative of volatility, are from Lodders (2003).
We find no evidence of the presence of Li \I\
$\lambda$6707, and report only an upper limit.
Entries are given for
several species that are obviously present, but for which
suitable weak lines were unavailable.  These cases are
indicated by `n.u,' for ``not used.''  Probable errors (pe)
are usually standard deviations of the abundances from `$n$' lines.  When
$n = 2,$ we used the difference between the values,
except when that difference was unrealistically small, in
which case a rough estimate is entered.  Solar abundances
are from Asplund, et al.(2009).  

Results of wavelength coincidence statistics or WCS
(Cowley \& Hensberge 1981) indicate that lines of 
Pr II and Gd II 
are present at 0.02 and 0.009 significance (or
false alarm probability) levels.
The certain presence of Ce II and Nd II leads us to 
believe Pr II and Gd II are indeed present, though
weak and blended.  We found no usable lines for an
abundance estimate.

Equivalent widths, oscillator strengths, and 
abundances are available in the online material
for some  318 absorption lines.

\begin{table}
\caption{Adopted abundances.  Column 2 is the condensation
temperature from Lodders (2003).  The number of lines
used is in the column labeled $n$.  Probable errors are
in Column 4.
\label{tab:abtab}}
\begin{minipage}{17cm}
\begin{tabular}{l r r c r r} \hline
Spec & $T_c$& $\log(El/\Sum)$&$\pm$pe&n    &Sun \\ \hline
Li {\sc i}&1135& $\le -10.95$&   & 1    \\
C {\sc i} &40  & $-$3.44 & 0.14  & 7   &$-$3.61\\
N {\sc i} &123 & $-$3.97 & 0.23  & 2   &$-$4.21 \\
O {\sc i} &179 & $-$3.31 & 0.07  & 5   &$-$3.35\\
Na {\sc i}&953 & $-$5.89 & 0.10  & 4   &$-$5.80\\
Mg {\sc i}&1327& $-$4.48 & 0.14  & 2   &$-$4.44\\
Mg {\sc ii}&   & n.u.    &       &         \\        
Al {\sc i}&1641& $-$5.92 & 0.20  & 2   &$-$5.59 \\
Si {\sc i}&1302& $-$4.69 & 0.30  & 7   &$-$4.52\\
Si {\sc ii}&    & n.u.   &       &       \\
S {\sc i}  &655 & $-$4.85 & 0.06  &  6 & $-$4.92 \\
Ca {\sc i} &1505& $-$6.00 & 0.16  &  7 &$-$5.70\\
Ca {\sc ii}&    & n.u.    &       &     \\
Sc {\sc ii}&1647& $-$9.18 & 0.11  &  5 & $-$8.98\\
Ti {\sc i} &    & $-$7.09 & 0.15  &  6   \\
Ti {\sc ii}&    & $-$7.21 & 0.08  &  3   \\
Ti         &1573& $-$7.17 & 0.15  &    &$-$7.09 \\
V {\sc i}  &    & $-$8.32 & 0.07  &  5            \\
V {\sc ii} &    & $-$8.39 & 0.14  &  3             \\
V          &1427& $-$8.32 & 0.07  &    &$-$8.11  \\
Cr {\sc i} &    & $-$6.52 & 0.31  & 12             \\
Cr {\sc ii}&    & $-$6.49 & 0.26  & 11              \\
Cr         &1291& $-$6.51 & 0.29  &    &$-$6.40    \\
Mn {\sc i} &1150& $-$6.95 & 0.19  & 15 &$-$6.61\\
Mn {\sc ii}&    & n.u.    &       &                \\
Fe {\sc i} &    & $-$4.79 & 0.08  & 10           \\
Fe {\sc ii}&    & $-$4.76 & 0.13  & 17      \\
Fe         &1328& $-$4.78 & 0.10  &    & $-$4.54  \\
Co {\sc i} &1347& $-$7.42 & 0.02  & 2  &$-$7.05 \\
Ni {\sc i} &1348& $-$5.98 & 0.12  & 16 &$-$5.82 \\
Ni {\sc ii}&    & n.u.    &       &  2           \\
Cu {\sc i} &1033& $-$8.12 & 0.34  &  2 & $-$7.84 \\
Zn {\sc i} &723 & $-$7.76 & 0.13  &  3 & $-$7.48\\
Sr {\sc ii}&1455& $-$9.45 & 0.1:  &  2 & $-$9.17\\
Y {\sc ii} &1647& $-$10.46& 0.12  &  6 & $-$9.83 \\
Zr {\sc ii}&1736& $-$9.73 & 0.21  & 4  & $-$9.46 \\
Ba {\sc ii}&1447& $-$10.27& 0.3:  & 2  & $-$9.86 \\ 
La {\sc ii}&1570& $-$11.24& 0.12  & 4  &$-$10.94 \\
Ce {\sc ii}&1477& $-$10.70 &0.14  & 5  &$-$10.46 \\
Nd {\sc ii}&1594& $-$10.77 &0.32  & 2  &$-$10.62 \\ \hline
\end{tabular}
\end{minipage}
\end{table}


\begin{figure}
\includegraphics[width=75mm,height=85mm,angle=-90]{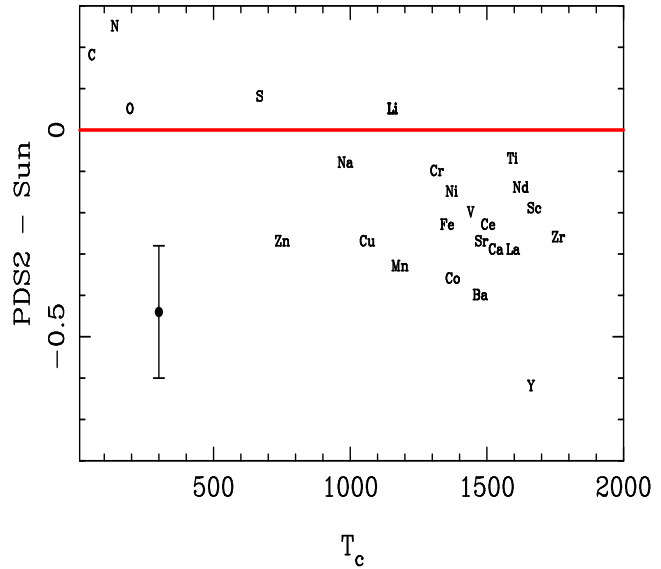}
 \caption{Logarithmic abundance differences between PDS2 and
the solar photosphere vs. condensation temperatures $T_C$ from Lodders(2003).  Data points are the lower left corner
of the chemical symbols.   An average error bar is shown at
the lower, left corner.  Individual values are in 
Table~\ref{tab:abtab}
\label{fig:DifTc}}
\end{figure}

Figure~\ref{fig:DifTc} shows a depletion of
volatile elements with respect to the Sun in PDS2.  Unlike
a similar result with HD 101412 (Paper I), the correlation of depletion with condensation temperature
appears monotonic--admittedly with considerable scatter.
Note that the intermediate volatile
zinc is depleted, unlike the general trend.
Folsom, et al. (2012) define a peculiarity index,
\begin{equation}
[P] =\frac{1}{3} ([Cr]+[Fe]+[Ni]-[C]-[N]-[O]),
\label{eq:P}
\end{equation}

\noindent which is a measure of the refractory 
depletion and the non-depletion (or excesses) of the
volatiles.

We find $[P] = -0.31$, which places PDS2 among the 
typical types of the Folsom, et al. study with negative $[P]$ indices.

\section{General remarks}

The similarity of the abundance peculiarities 
of some Herbig Ae and
those of the $\lambda$ Boo stars is becoming firm.  We
have now studied 4 Herbig Ae stars with relatively 
sharp lines.  Two have shown the noted peculiarities, 
while two have not.  These results match those of the
broader survey of Folsom, et al (2012), who concluded
``half the stars in our sample show $\lambda$ Boo 
chemical peculiarities to varying degrees.''  Their work
included HD 101412 and HD 190073.  They found the 
former star to have $\lambda$ Boo-like,
and the later to have solar abundances, as we
reported in Papers I and II.

The connection between young stars and $\lambda$ Boo
abundances was noted by Gray and Corbally (1998),
whose monograph (Gray and Corbally 2009) contains
a description of $\lambda$ Boo and Herbig 
Ae/Be stars, as well as theories of their origin.
These authors and others (e.g. Heiter, et al. 2002)
note that this general pattern--the selective depletion
of refractory elements--occurs in certain RV Tauri and
post-AGB stars, objects with vastly different structure
and evolutionary history from Herbig Ae stars. 

More recently (cf. Mel\'{e}ndez 2013), precision abundances
of the Sun and solar-like stars have revealed
a similar pattern of depletion of refractory elements in 
the Sun itself.

The most common explanations of how these abundance
patterns arise have built on
the suggestions of
Venn and Lambert (1990).  They noted the similarity of the
$\lambda$ Boo abundances to those of the interstellar gas
from which refractory elements have been lost due to
depletion onto grains.  
The grain-gas separation could take place in interstellar or
circumstellar environments.  Mel\'{e}ndez (2013) has suggested
the sun's depletion of refractory elements is due to 
the formation of (metal/rocky) terrestrial planets.
Quite different models may be required for the diverse
objects that share the pattern of refractory-element 
depletions.  But all are likely to draw upon the 
chemo-thermodynamic properties of the elements.

The Herbig Ae stars, must represent the progenitors of
normal A stars. However, since half of these young
A-stars are chemically peculiar, some explanation of how
the older stars become normal or metal rich is required.
The ages of $\lambda$ Boo stars are quite uncertain
due to their near absence in galactic clusters.  The
Herbig Ae stars, however, are definitely young.  
It is plausible that their abundance anomalies can be 
destroyed by diffusion or meridonal mixing, as 
suggested by Turcotte (2002).


\section{Acknowledgements}

We thank J. R. Fuhr J. Reader, and W. Wiese of NIST for
advice on atomic data and processes, and J. F.
Gonz\'{a}lez for help with the observational material.
This research has made use of the SIMBAD database,
operated at CDS, Strasbourg, France. Our calculations
have made extensive use of the 
VALD\footnote{http://vald.astro.univie.ac.at/$\sim$vald/php/vald.php}
atomic data base
(Kupka, et al. 1999), as well as the 
NIST\footnote{http://www.nist.gov/pml/data/index.cfm}
online Atomic
Spectroscopy Data Center (Kramida, et al. 2013). CRC is
grateful for advice and helpful conversations with many
of his Michigan colleagues.   He thanks
Jesus Hern\'{a}ndez for a discussion of the 
status of PDS2, and C. Folsom for a useful email exchange
on the possible chemical evolution of Herbig Ae stars.

This research is based archival data from 
observations
obtained at the European Southern
Observatory, Paranal and La Silla, Chile
(ESO programmes 082.D-0833(A),
Archive request Nos. 54506 SAF
and 59603 SAF).

\section*{REFERENCES}





\hangpar Asplund, M., Grevesse, N., Sauval, A. J., Scott, P.
2009, Ann. Rev. Astron. Ap. 47, 481

\hangpar Bagnulo, S., Landstreet, J. D., Fossati, L. \&
Kochukhov, O. 2012, A\&A, 538, 129



\hangpar Barklem, P.S. 2007, A\&A, 462, 781

\hangpar Barklem, P.S., Piskunov, N., O'Mara, B.J. 2000, A\&AS, 142, 467 

\hangpar Bernabei, S., Marconi, M., Ripepi, V., Leccia, S.,
Rodr\'{i}guez, E., Oswalt, T. D., et al. 2007, Comm. in
Asteroseismology, 150, 2007






\hangpar Boyajian, T. S., von Braun, K., van Belle, G., 
Farrington, C., Schaefer, G., Jones, J., et al. 2013,
ApJ, 771, 40

\hangpar Brault, J. W. \& White, O. R. 1971, A\&A, 13, 169

\hangpar Butler, K., \& Giddings, J.R.~1985, in Newsletter of Analysis of
Astronomical Spectra, No.~9 (Univ.~London)



\hangpar Castelli, F. \& Kurucz, R. L. 2003, in
Modelling of Stellar Atmospheres, ed. N. Piskunov,
W. W. Weiss, D. F. Gray, IAU Symposium 210,
(A 20) p. 424

\hangpar Cowley, C. R., Adelman, S. J. \& Bord, D. J. 2003,
in Modelling of Stellar Atmospheres, ed. N. Piskunov, W. W.
Weiss, \& D. F. Gray (IAU Symposium 210), p. 261

\hangpar Cowley, C. R. \& Hensberge, H. 1981, ApJ, 244, 252

\hangpar Cowley, C. R., Hubrig, S., Gonz\'{a}lez \&
Savanov, I. 2010, A\&A, 523, 65 (Paper I)

\hangpar Cowley, C. R. \& Hubrig, S. 2012, AN, 333, 34 (Paper II)

\hangpar  Cowley, C. R., Hubrig, S., Castelli, F. \& Wolff, B.  2012, A\&A, 537, L6



\hangpar Cowley, C. R., Castelli, F. \& Hubrig, S. 2013,
MNRAS, 431, 3485 (Paper III)













\hangpar Folsom, C. P., Bagnulo, S., Wade, G. A., Alecian, E., Landstreet, J. D., Marsden, S. C. \& Wate, I. A. 2012, 
MNRAS, 422, 2072

\hangpar Folsom, C. P., Bagnulo, S., Wade, G. A., Landstreet, J. D. \& Alecian, E. 2013, Magnetic fields 
throughout stellar evolution, Proc. IAU Symp. 302, ed.
P. Petit, arXiv:1311.1552

\hangpar Froese Fischer, C., \& Tachiev, G. 2004, 
At. Data Nucl. Data Tables, 87, 1




\hangpar Giddings, J.R.~1981, Ph.D.~Thesis (Univ.~London)


\hangpar Gray, D. F. 2005, The Observation and Analysis
of Stellar Photospheres, 3rd ed. (Cambridge: University Press,
see p. 322)

\hangpar Gray, R. O., \& Corbally, C. J. 1998, AJ,
116, 2530

\hangpar Gray, R. O. \& Corbally, C. J. 2009, Stellar
Spectral Classification (Princeton: Series in Astrophys.),
see pp. 199--203

\hangpar Gregorio-Hetem, J., L\'{e}pine, J. R. D., Quast, G.
R., Torres, C. A. O. \& de la Reza, R. 1992, AJ, 103, 549





\hangpar Heiter, U., Weiss, W. W. \& Paunzen, E. 2002, A\&A,
381, 971





\hangpar Hansen, C. J., Bergemann, M., Cescutti, G., Francois, P.,
Arcones, A., Karakas, A. I, et al. 2013, A\&A, 551, 57

\hangpar Hubrig, S., Stelzer, B., Sch\"{o}ller, M.,
Grady, C., Sch\"{u}tz, Pogodin, M. A., et al. 2009,
A\&A, 502, 283

\hangpar Hubrig, S., Illyn, I., Cowley, C. R., Castelli, F., 
Stelzer, B., et al. 2013, arXiv:1308.6777 (to appear in
Proceedings of the Conference ``Physics at the Magnetospheric
Boundary'', EDP Sciences, in press.




\hangpar IRAS Science Team, 1988, Catalogs and Atlases, 
Vol. 2--6: The Point Source Catalog (PSC), ed. Beichman,
C. A., Neugebauer, G., Habing, H. J., Clegg, P. E. \&
Chester, T. J., NASA RP-1190


\hangpar K\"{a}ufl, H.-U., Ballester, P., Biereichel, P.,
Delabre, B., Donaldson, R., Dorn, R., et al. 2004, SPIE,
5492, 1218

\hangpar Kramida, A., Ralchenko, Yu., Reader, J., and NIST ASD Team (2013). NIST Atomic Spectra Database (ver. 5.0), [Online]. Available: http://physics.nist.gov/asd [2013, February 21]. National Institute of Standards and Technology, Gaithersburg, MD. 

\hangpar Kupka, F., Piskunov, N. E., Ryabchikova, T. A.,
Stempels, H. C., Weiss, W. W. 1999, A\&AS, 138, 119

\hangpar Kurucz, R. L. 1993, ATLAS9 Stellar Atmosphere
Programs and 2 km/s grid, CD-Rom, No. 13, Cambridge MA:
Smithsonian Ap. Obs.


\hangpar Kurucz, R. L. 1996, in Model atmospheres and 
spectrum synthesis, ASP Conf. Ser. 108, ed. S. J. Adelman,
F. Kupka \& W. W. Weiss, p. 160








\hangpar Liebert, J., Fontaine, G., Young, P. A., Williams,
K. A. \& Arnett, D. 2013, ApJ, 769, 7

\hangpar Lodders, K. 2003, ApJ, 591, 1220






\hangpar Marconi, M., Ripepi, V., Bernabei, S.,
Ruoppo, A., Monteiro, M. J. P. F. G.,Marques, J. P.,
et al. Astrophys. Sp. Sci. 2010, 328, 109

\hangpar Mashonkina, L., Gehren, T., Shi, J.-R., Korn,
A. J. \& Grupp, F. 2011, A\&A, 528, 87


\hangpar Mayor, M., Pepe, F., Queloz, D., Bouchy, F.,
Rupprecht, G., Lo Curto, G., et al. 2003, The Messenger, 114, 20



\hangpar Mel\'{e}ndez, J. 2013, arXiv:1307.5274 (to appear
in Setting the scene for Gaia and LAMOST, Proc. IAU 
Symposium No. 298)










\hangpar Pogodin, M. A., Hubrig, S., Yudin, R. V., Sch\"{o}ller,
M., Gonz\'{a}lez, J. F. \& Stelzer, B. 2012, AN, 333, 594

\hangpar Przybilla, N. \& Butler, K. 2004, ApJ,
610, L61

\hangpar Przybilla, N., Butler, K., Becker, S.R., Kudritzki, R.P., Venn, K.A. 2000, A\&A, 359, 1085







\hangpar Steenbock, W. \& Holweger, H. 1984, A\&A, 130, 319

\hangpar Turcotte, S. 2002, ApJ, L129





\hangpar Venn, K. A., Lambert, D. L. 1990, ApJ, 363, 234

\hangpar Vernet, J., Dekker, H., D'Odorico, S. D., Kaper, L.,
Kjaergaard, P., Hammer, F., et al. 2011, A\&A 


\hangpar Vieira, S. L. A., Corradi, W. J. B., Alencar,
S. H. P., Mendes, L. T. S., Torres, C. A. O.,
Quast, G. R., Guimar\~{a}es, M. M., da Silva, L. 2003,
AJ, 126, 2971

\hangpar Wade, G. A., Bagnulo, S., Drouin, D., Landstreet, 
J. D. \& Monin, D. 2007, MNRAS, 376, 1145


\hangpar Wiese, W.L., Fuhr, J.R., \& Deters, T.M. 1996, 
J. Phys. \& Chem. Ref. Data, Mon. 7

\appendix
\section{Online material}

The online material gives wavelengths, measured
equivalent widths, oscillator strengths, and lower
excitation potentials for individual lines.  References
to original papers may be found in Tables 4 and A1 of
Paper III.  The online file is plain ascii and machine
readable. The subset of weaker lines used for abundances
are marked with an asterisk. The stronger lines were
used for the microturbulence determination.  Additional
information is provided on the choice of oscillator
strengths, and the relation of the adopted to the solar
abundance.


\begin{table}
\end{table}
\begin{verbatim}
Table A1: Sample of online material.
Sodium
Na I - Log(Na/Ntot) = -5.89 +/- 0.1.  Results from
the 4 weakest lines adopted.  The solar value
is -5.80.  Sodium is solar to within the errors.

Wavelen   W[mA]  log(W) log(gf)  Chi(eV) LogN/Ntot
--------------------------------------------------
4497.657* 12.9    1.11  -1.560   2.100   -5.99
4668.559* 29.5    1.47  -2.250   2.100   -5.88
5682.633  82.8    1.92  -0.700   2.100   -5.63
5688.205 110.0    2.04  -1.400   2.100   -5.56
5889.951 357.0    2.55   0.108   0.000   -5.64
5895.924 308.0    2.49  -0.194   0.000   -5.53
6154.226* 22.7    1.36  -1.547   2.100   -5.76
6160.747* 28.5    1.45  -1.246   2.100   -5.93
\end{verbatim}
\label{lastpage}
\end{document}